\documentclass[twocolumn,showpacs,preprintnumbers,amsmath,amssymb]{revtex4}
\usepackage{graphicx}
\usepackage{dcolumn}
\usepackage{bm}
\begin{document}
\title{Spin susceptibility and polarization field in a dilute two-dimensional electron system in (111) silicon}
\author{A.~A. Kapustin, A.~A. Shashkin, and V.~T. Dolgopolov}
\affiliation{Institute of Solid State Physics, Chernogolovka, Moscow District 142432, Russia}
\author{M. Goiran and H. Rakoto$^*$}
\affiliation{CNRS, LNCMP, 143 Avenue de Rangueil, F-31400 Toulouse, France\\
Universite de Toulouse, UPS, INSA, LNCMP, F-31077 Toulouse, France}
\author{Z.~D. Kvon}
\affiliation{Institute of Semiconductor Physics, Novosibirsk, 630090, Russia}
\begin{abstract}
We find that the polarization field, $B_\chi$, obtained by scaling
the weak-parallel-field magnetoresistance at different electron
densities in a dilute two-dimensional electron system in (111)
silicon, corresponds to the spin susceptibility that grows strongly
at low densities. The polarization field, $B_{\text{sat}}$,
determined by resistance saturation, turns out to deviate to lower
values than $B_\chi$ with increasing electron density, which can be
explained by filling of the upper electron subbands in the fully
spin-polarized regime.
\end{abstract}
\maketitle

Much interest has been attracted recently to the anomalous properties
of strongly interacting two-dimensional (2D) electron systems. The
interaction strength is characterized by the Wigner-Seitz radius,
$r_s=1/(\pi n_s)^{1/2}a_B$ (where $n_s$ is the electron density and
$a_B$ is the effective Bohr radius in semiconductor), which is equal
in the single-valley case to the ratio of the Coulomb and the Fermi
energies. While at high electron densities ($r_s<1$) conventional
Fermi-liquid behavior is established, at very low electron densities
($r_s\gg1$) formation of the Wigner crystal is expected
\cite{chaplik72,tanatar89,attaccalite02}. It was not until recently
that qualitative deviations from the weakly-interacting Fermi liquid
behavior (in particular, the drastic increase of the effective
electron mass with decreasing electron density) have been found in
strongly correlated 2D electron systems ($r_s\gtrsim10$)
\cite{review,shashkin06,anissimova06}.

The strongest many-body effects have been observed in (100)-silicon
metal-oxide-semiconductor field-effect transistors (MOSFETs). This
has particularly stimulated a new series of experiments in
(111)-silicon MOSFETs \cite{estibals04,eng05,eng07,shashkin07}. It
was found in Ref.~\cite{shashkin07} by an analysis of the
temperature-dependent Shubnikov-de~Haas oscillations that the
strongly increased mass is also the case in a dilute 2D electron
system in (111) silicon with relatively high disorder, as indicated
by both the absence of metallic temperature dependence of zero-field
resistance and the relatively low mobility. Moreover, it was found
that the relative mass enhancement as a function of $r_s$ is system-
and disorder-independent being determined by electron-electron
interactions only. However, no sign of the enhanced spin
susceptibility, $\chi$, which is proportional to the product of the
$g$ factor and effective mass, $m$, was observed in parallel-field
magnetoresistance measurements \cite{estibals04}. Rather than tending
to zero at a finite electron density like in (100)-silicon MOSFETs
\cite{review}, the resistance saturation field, $B_{\text{sat}}$,
corresponding to the onset of full polarization of the electrons'
spins was found to increase approximately proportionally with the
electron density. It is worth noting that the valley degeneracy in
(111)-silicon MOSFETs is equal to $g_v=2$ (as well as the spin
degeneracy $g_s=2$), rather than $g_v=6$ (see, e.g.,
Ref.~\cite{neugebauer75}), as inferred from the filling factor change
for the period of Shubnikov-de~Haas oscillations in weak magnetic
fields, equal to 4. The reduced valley degeneracy was attributed, in
particular, to the existence of inhomogeneous strains at the
interface \cite{tsui76}. The upper electron subbands can be important
for the behavior of $B_{\text{sat}}$. Filling of these subbands has
been observed recently in studies of Shubnikov-de~Haas oscillations
in (111)-silicon samples of a special design: a 2D electron system
forms at the interface between (111) silicon and vacuum \cite{eng07}.

In this paper, we study low-temperature parallel-field
magnetotransport in a 2D electron system in (111) silicon in a wide
range of electron densities. We find that the polarization magnetic
field, $B_\chi$, obtained by scaling the weak-parallel-field
magnetoresistance at different electron densities, tends to zero at a
finite electron density in a linear fashion. It corresponds to the
spin susceptibility $\chi\propto gm$ that increases strongly at low
densities, $B_\chi\propto n_s/\chi$, which is consistent with the
increase of $m$ observed earlier in this electron system. The
polarization field $B_{\text{sat}}$, determined by resistance
saturation, turns out to deviate to lower values than $B_\chi$ with
increasing electron density. This can be explained by the
parallel-field-induced filling of the upper electron subbands
corresponding to another two valleys in (111) silicon. The subbands'
splitting in our samples is estimated at $\Delta\approx2$~meV.

Measurements were made in an Oxford dilution refrigerator with a base
temperature of $\approx 30$~mK (Chernogolovka) and a dilution
refrigerator with a base temperature of $\approx 80$~mK (Toulouse) in
magnetic fields up to 14~T on (111)-silicon MOSFETs similar to those
previously used in Ref.~\cite{estibals04}. Samples had the Hall bar
geometry with width 400~$\mu$m equal to the distance between the
potential probes. Application of a dc voltage to the gate relative to
the contacts allowed one to control the electron density. Oxide
thickness was equal to 154~nm. In highest-mobility samples, the
normal of the sample surface was tilted from [111]- toward
[110]-direction by a small angle of $8^\circ$. Anisotropy for
electron transport in such samples does not exceed 5\% at
$n_s=3\times10^{11}$~cm$^{-2}$ and increases weakly with electron
density, staying below 25\% at $n_s=3\times10^{12}$~cm$^{-2}$
\cite{shashkin07}. The resistance was measured by a standard
4-terminal technique at a low frequency (5--10~Hz) to minimize the
out-of-phase signal. Excitation current was kept low enough (below
2~nA) to ensure that measurements were taken in the linear regime of
response. The current was parallel to the in-plane magnetic field.
Electron densities were determined by Shubnikov-de~Haas oscillations
in perpendicular magnetic fields, for which the sample orientation
was changed at room temperature and another run was made. The
threshold voltage was checked to be the same in different runs,
although the peak electron mobility at $T\approx1.5$~K varied in the
range between 2500 and 4000~cm$^2$/Vs. Two samples were used in the
experiments; below, we show data obtained on one of them. The results
reported are independent of a sample and a sample state and are
reproducible in different cryostats.

Additional measurements were made in a rotator-equipped He$^4$
cryostat at a temperature of $\approx 1.5$~K in pulsed magnetic
fields (Toulouse). In each pulse, the magnetic field swept up to
$\approx 48$~T (or lower) with rising and falling times of about
50~ms and 0.3~s, respectively. Excitation current with frequency
2~kHz did not exceed 600~nA.

\begin{figure}
\scalebox{0.51}{\includegraphics{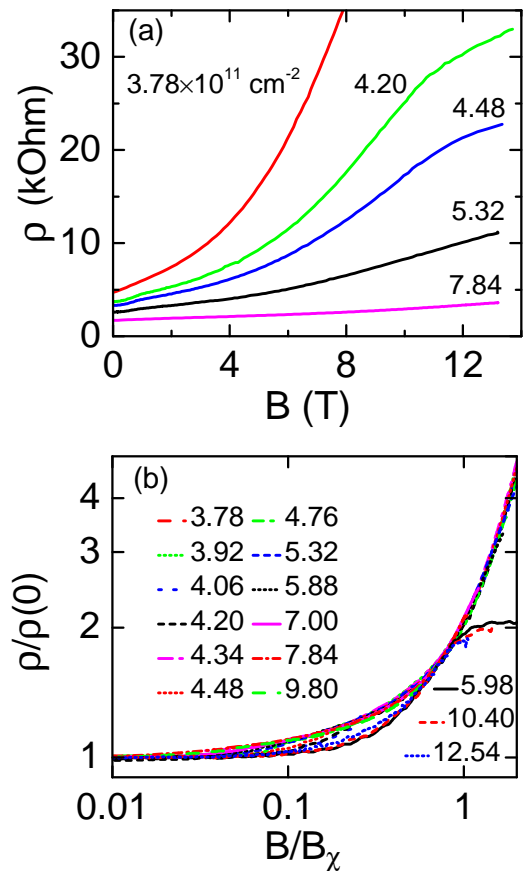}}
\caption{\label{fig1} (a)~Magnetoresistance in dc parallel magnetic
fields at a temperature of $\approx80$~mK at different electron
densities in the metallic regime. (b)~Scaled curves of the normalized
magnetoresistance at different $n_s$ versus $B/B_\chi$ for dc (upper
set) and pulsed (lower set) parallel magnetic fields. The electron
densities are indicated in units of $10^{11}$~cm$^{-2}$.}
\end{figure}

\begin{figure}
\scalebox{0.48}{\includegraphics{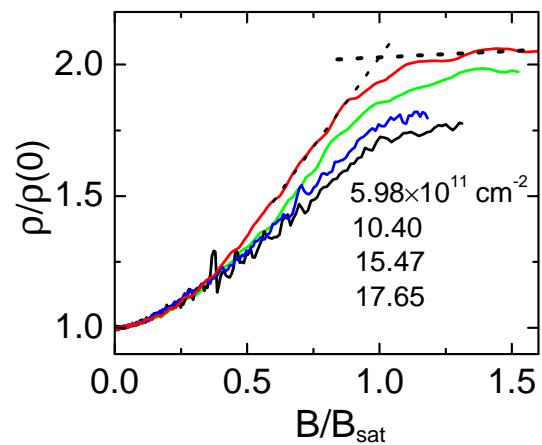}}
\caption{\label{fig2} The normalized magnetoresistance as a function
of $B/B_{\text{sat}}$ at different electron densities for pulsed
parallel magnetic fields. Also shown is the way to determine the
saturation field.}
\end{figure}

Typical curves of the low-temperature magnetoresistance $\rho(B)$ in
a parallel magnetic field are displayed in Fig.~\ref{fig1}(a). The
resistivity increases with field and saturates above a certain
density-dependent magnetic field. Deep in the metallic regime
($\rho\ll h/e^2$), the magnetoresistance depends weakly on
temperature in the experimental range of temperatures (at
$T\leq1.5$~K). This limit is realized for progressively narrower
initial intervals on the curve $\rho(B)$, as the electron density is
lowered. The data discussed in this paper are obtained deep in the
metallic regime.

In Fig.~\ref{fig1}(b), we show how the normalized weak-field
magnetoresistance, measured at different electron densities,
collapses onto a single curve when plotted as a function of
$B/B_\chi$. The scaling parameter $B_\chi$ has been normalized to
correspond to the magnetic field at which the magnetoresistance at
the lowest electron densities in the good metallic regime saturates
and full spin polarization of the electrons is reached
\cite{okamoto99,vitkalov00,shashkin01}. Within the accuracy with
which the saturation field $B_{\text{sat}}$ can be determined, we use
for the normalization the curve $\rho(B)$ at $n_s\approx6\times
10^{11}$~cm$^{-2}$, measured in pulsed magnetic fields; we have
verified that the fields $B_{\text{sat}}$ and $B_\chi$ are
practically temperature-independent in the range of temperatures
used. At high electron densities, the value of $B_{\text{sat}}$ is
noticeably smaller than $B_\chi$ in this electron system, as is
evident from Fig.~\ref{fig2} which shows the normalized
magnetoresistance versus $B/B_{\text{sat}}$ in pulsed magnetic
fields. Apparently, the scaled data in Fig.~\ref{fig1}(b) are a
function of the degree of spin polarization, $\xi=B/B_\chi$, in weak
magnetic fields but not at $\xi\approx1$.

\begin{figure}
\scalebox{0.48}{\includegraphics{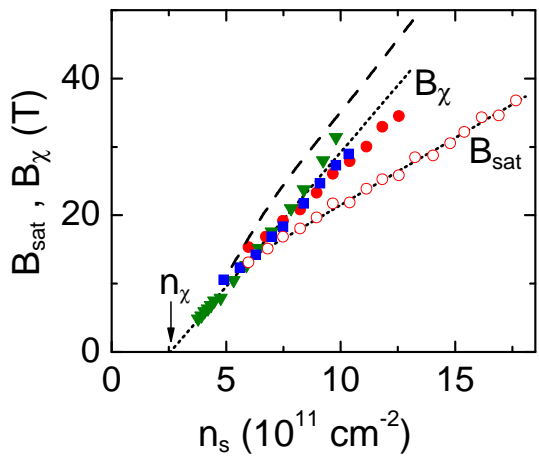}}
\caption{\label{fig3} Dependence of the fields $B_{\text{sat}}$ (open
circles) and $B_\chi$ (squares, triangles, and dots) on electron
density. Squares and triangles are obtained in two runs in dc
magnetic fields, and dots and open circles correspond to pulsed
magnetic fields. The dotted lines are linear fits. The fit for
$B_\chi$ extrapolates to a critical density $n_\chi$. Also shown by a
dashed line is the data for the enhanced effective mass of
Ref.~\cite{shashkin07} recalculated into $B_\chi$ using the value of
$g=2$ in bulk silicon.}
\end{figure}

In Fig.~\ref{fig3}, we show the so-determined polarization fields
$B_{\text{sat}}$ and $B_\chi$ as a function of electron density. The
field $B_\chi$ obtained by scaling the weak-field magnetoresistance
depends linearly on $n_s$ and tends to vanish at a finite electron
density, $n_\chi$. Our procedure provides good accuracy for
determining the behavior of $B_\chi$ with electron density, i.e., the
functional form of $B_\chi(n_s)$, even though the absolute value of
$B_\chi$ is determined not so accurately. Based on the form of the
spin polarization parameter $\xi=\chi
B/\mu_Bn_s=g\mu_BB/2E_F=B/B_\chi$ (where $\mu_B$ is the Bohr magneton
and $E_F=\pi\hbar^2n_s/2m$ is the Fermi energy in $B=0$), we have
recalculated the data for the enhanced effective mass $m(n_s)$ of
Ref.~\cite{shashkin07} into $B_\chi(n_s)$ (dashed line), using the
value of $g=2$ in bulk silicon which is close to the $g$ factor
measured in silicon MOSFETs \cite{review}. Both dependences
$B_\chi(n_s)$ are in agreement with each other, and the spin
susceptibility $\chi\propto gm$ is strongly enhanced at low
densities. The resistance saturation field $B_{\text{sat}}$ deviates
to lower values than $B_\chi$ with increasing electron density,
$B_{\text{sat}}(n_s)$ being a linear dependence with approximately
twice as small a slope. We note that our data for $B_{\text{sat}}$
are concurrent with earlier results \cite{estibals04}.

The observed behavior of the polarization field $B_\chi$ with
electron density, which corresponds to the sharply increased spin
susceptibility at low densities, is very similar to that in a 2D
electron system in (100)-silicon MOSFETs \cite{shashkin01}. The same
interaction strength in (111)-silicon MOSFETs is expected to be
achieved at electron densities about four times higher than in (100)
silicon, since in the first case the effective mass $m_b=0.358m_e$
(where $m_e$ is the free electron mass) in bulk silicon is
approximately twice as large. Based on the critical density,
$n_\chi\approx8\times10^{10}$~cm$^{-2}$, of vanishing polarization
field for (100)-silicon MOSFETs, one expects the critical region in
(111) silicon at electron densities about $3\times10^{11}$~cm$^{-2}$.
This is consistent with our results for $B_\chi$ (Fig.~\ref{fig3}).

It is interesting that the polarization field $B_{\text{sat}}(n_s)$
behaves differently from $B_\chi(n_s)$. This finding is in contrast
to the case of (100)-silicon MOSFETs where the fields $B_\chi$ and
$B_{\text{sat}}$ are practically the same \cite{review} and the
scaled data are described reasonably well by the theoretical
dependence of the normalized magnetoresistance on the degree of spin
polarization \cite{dolgopolov00} in a wide range of $\xi$. Clearly,
the discrepancy between both polarization fields in (111)-silicon
MOSFETs reflects the fact that the spin polarization parameter is not
linear in $B$ at large $\xi\approx1$. Below, we discuss possible
mechanisms for the effect.

A simple reason for the value of saturation field to become reduced
as the electron density is increased might be the orbital effects
that occur when the thickness of the 2D electron system becomes
comparable to the magnetic length, $l_B=(\hbar c/eB)^{1/2}$. These
are expected to lead to an increase of the effective mass with
parallel magnetic field (see, e.g., Ref.~\cite{tutuc03}). However, no
such reduction in the $B_{\text{sat}}(n_s)$ dependence has been found
for (100)-silicon MOSFETs in the same range of pulsed parallel
magnetic fields \cite{broto03}. Because the corresponding thicknesses
of the 2D electron systems \cite{ando82} for both orientations in
silicon MOSFETs are close to each other, the orbital effects are
likely to be small for our case.

\begin{figure}
\scalebox{0.48}{\includegraphics{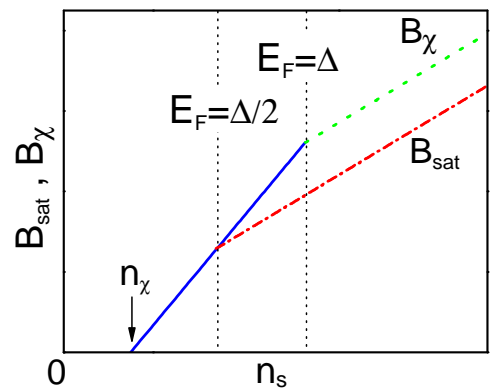}}
\caption{\label{fig4} The schematic behavior of the fields
$B_{\text{sat}}$ and $B_\chi$ with electron density, as discussed in
text. There are break points in the dependences $B_{\text{sat}}(n_s)$
at $E_F=\Delta/2$ (or $E_F(B_{\text{sat}})=\Delta$) and $B_\chi(n_s)$
at $E_F=\Delta$.}
\end{figure}

Another reason for the observed reduction of $B_{\text{sat}}$ is the
filling of the upper electron subbands in the fully spin-polarized
regime. The Fermi energy should double at the onset of complete spin
polarization \cite{dolgopolov00}, if $E_F$ is below the mid-gap,
$\Delta/2$ (here $\Delta$ is the subband separation). In the opposite
case $E_F>\Delta/2$ the parallel magnetic field promotes filling of
the upper subbands. These are empty in the experimental range of
$n_s$ in our samples in $B=0$ (i.e., $E_F<\Delta$) but can be filled
at high electron densities in parallel magnetic fields provided the
condition $E_F>\Delta/2$ is met. In this case the Fermi energy at the
onset of complete spin polarization,
$E_F(B_{\text{sat}})=g\mu_BB_{\text{sat}}$, exceeds the splitting
$\Delta$ and should increase with electron density with a slope
reduced by the factor of two (see Fig.~\ref{fig4}), as can be
expected from the results of Ref.~\cite{eng07}. There, the filling
factor change for the period of the weak-field Shubnikov-de~Haas
oscillations in (111)-silicon samples has been found to be equal to
8, which immediately indicates that the upper subbands in question
correspond to another two valleys. The slope reduction expected for
both $E_F(B_{\text{sat}})$ and $B_{\text{sat}}$ is in agreement with
the data in Fig.~\ref{fig3}. Note that the observed linear increase
of $B_{\text{sat}}$ with $n_s$ indicates that a possible influence of
the diamagnetic contribution to $\Delta$ is small. Thus, the
parallel-field-induced filling of the upper subbands corresponding to
another two valleys in (111) silicon can account for the experimental
behavior of the saturation field.

We estimate the splitting of the subbands at $\Delta\approx2$~meV for
$n_s\approx6\times10^{11}$~cm$^{-2}$. This value is appreciably
bigger than the splitting $\Delta\approx0.6$~meV obtained in
Ref.~\cite{eng07}, which is consistent with the pattern of
Shubnikov-de~Haas oscillations in the two sets of samples. The fact
that the splitting $\Delta$ is not universal in different samples may
be due to different inhomogeneous strains at the interface
\cite{tsui76} which are sensitive to a particular procedure of sample
fabrication. Based on the experimental data and ignoring the possible
(linear) dependence of $\Delta$ on $n_s$, one can expect filling of
the upper subbands in $B=0$ at electron densities above
$\approx1.5\times10^{12}$~cm$^{-2}$. Once no double the periodicity
of Shubnikov-de~Haas oscillations in weak magnetic fields was
observed in our samples up to $n_s\approx2.5\times10^{12}$~cm$^{-2}$
\cite{shashkin07}, yet higher electron densities should be tried to
clarify the point. Note that at very high electron densities about
$1\times10^{13}$~cm$^{-2}$, observation of the sixfold valley
degeneracy was reported \cite{tsui79}, although the high $g_v$ can be
promoted by additional annealing of the samples \cite{cole84}.

In summary, we have found that the polarization field $B_\chi$,
obtained by scaling the weak-parallel-field magnetoresistance at
different electron densities in (111)-silicon MOSFETs, tends to zero
at a finite electron density in a linear fashion. It corresponds to
the spin susceptibility that increases strongly at low densities.
This is consistent with the increase of the effective mass observed
earlier in this electron system. The polarization field
$B_{\text{sat}}$, determined by resistance saturation, turns out to
deviate to lower values than $B_\chi$ with increasing electron
density. The reduction of $B_{\text{sat}}$ can be explained by
filling of the upper electron subbands in the fully spin-polarized
regime. The results obtained allow us to estimate the subbands'
splitting in our samples.

We gratefully acknowledge discussions with A. Gold and S.~V.
Kravchenko. We would like to thank M. Nardonne for technical
assistance. This work was supported by the RFBR, RAS, and the
Programme ``The State Support of Leading Scientific Schools''.

\end{document}